\documentclass[11pt]{article}

\newcommand{\cmp}{Comm. Math. Phys.~}

\newcommand{\jmp}{J. Math. Phys.~}

\newcommand{\njp}{New. J. Phys.~}

\newcommand{\prl}{Phys. Rev. Lett.~}
\newcommand{\pra}{Phys. Rev. A~}
\newcommand{\pre}{Phys. Rev. E~}

\usepackage{young}
\usepackage{color}
\usepackage[latin1]{inputenc}
\usepackage{appendix}
\usepackage{epstopdf}
\usepackage{color}
\usepackage{subfigure}
\usepackage[T1]{fontenc}
\usepackage[sc]{mathpazo}
\usepackage{amsmath}
\usepackage{amssymb}
\usepackage{graphicx,color}
\usepackage{framed}
\usepackage{multirow}
\usepackage{enumerate}
\usepackage{amsthm}
\usepackage{booktabs}
\usepackage{amsfonts,mathrsfs}
\usepackage{geometry} 
\usepackage{eepic}
\usepackage{ifthen}
\usepackage[vcentermath]{youngtab}
\usepackage[unicode=true,pdfusetitle, bookmarks=true,bookmarksnumbered=false,bookmarksopen=false, breaklinks=false,pdfborder={0 0 0},backref=false,colorlinks=false] {hyperref}
\hypersetup{
colorlinks,linkcolor=myurlcolor,citecolor=myurlcolor,urlcolor=myurlcolor}
\definecolor{myurlcolor}{rgb}{0,0,0.7}

\newcommand{\blue}{\textcolor{blue}}

\usepackage{hyperref}
\hypersetup{pdfpagemode=UseNone}

\def\Span{\mathrm{span}}

\def \dif {\mathrm{d}}

\newenvironment{mylist}[1]{\begin{list}{}{
    \setlength{\leftmargin}{#1}
    \setlength{\rightmargin}{0mm}
    \setlength{\labelsep}{2mm}
    \setlength{\labelwidth}{8mm}
    \setlength{\itemsep}{0mm}}}
    {\end{list}}

\newcommand{\Pa}[1]{\left(#1\right)}

\newcommand{\Set}[1]{\left\{#1\right\}}

\newcommand{\bra}[1]{\langle#1|}

\newcommand{\ket}[1]{|#1\rangle}

\DeclareMathOperator{\trace}{Tr}

\newcommand{\Ptr}[2]{\trace_{#1}\Pa{#2}}

\newcommand{\Tr}[1]{\Ptr{}{#1}}

\def\cL{\mathcal{L}}

\def\bsA{\boldsymbol{A}}\def\bsB{\boldsymbol{B}}
\def\bsF{\boldsymbol{F}}\def\bsH{\boldsymbol{H}}\def\bsI{\boldsymbol{I}}
\def\bsL{\boldsymbol{L}}
\def\bsP{\boldsymbol{P}}
\def\bsU{\boldsymbol{U}}

\def\bsf{\boldsymbol{f}}\def\bsj{\boldsymbol{j}}

\def\bsp{\boldsymbol{p}}
\def\bsx{\boldsymbol{x}}

\theoremstyle{definition}

\newcounter{questionnumber}

\usepackage[font=footnotesize]{caption}
\usepackage{etoolbox}

\AtBeginEnvironment{table}{\small}
\AtBeginEnvironment{table*}{\small}
\AtBeginEnvironment{figure}{\footnotesize}
\AtBeginEnvironment{figure*}{\footnotesize}

\begin{document}

\title{\bf \large Quantum Probability Current Guided Reduction of Coupling Control Degrees of Freedom for Excitation Transport}
\author{\blue{Liuheng Cao}$^1$\footnote{Email:
liuhengcao@163.com},\quad \blue{Lin Zhang}$^1$\footnote{Email:
godyalin@163.com (corresponding author)},\quad \blue{Junde
Wu}$^2$\footnote{Email: wjd@zju.edu.cn}
\\
  {\it\small $^1$School of Mathematical Sciences, Hangzhou Dianzi University, Hangzhou 310018, China}\\
  {\it\small $^2$School of Mathematical Sciences, Zhejiang University, Hangzhou 310027, China}
  }
\date{}
\hypersetup{pdfauthor={Liuheng Cao, Lin Zhang, and Junde Wu}}
\maketitle

\begin{abstract}
Time-dependent coherent control can enhance excitation transport in open
quantum networks, but independently controlling every inter-site coupling
creates a control space of high dimension and leads to difficult optimization problems. We introduce an edge-ranking strategy based on the control-induced change in the gradient component of the time-integrated quantum probability current, which is obtained via a graph Hodge decomposition. When our strategy is applied to the seven-site Fenna-Matthews-Olson (FMO) model, the six-edge set retains $99.83\%$ of the enhancement achieved by full control, and the four-edge set retains $97.60\%$ while reducing the pulse fluence---used here as a proxy for control effort---by $41.55\%$ relative to full control. Dephasing scans and comparisons with random edge sets and random networks provide numerical support for the relevance and potential broader utility of the ranking. These results show that edge selection guided by the quantum probability current can substantially reduce the control space while preserving high transport performance with lower control effort.
\end{abstract}

\section{Introduction}

Quantum effects in biological systems have attracted considerable
attention, particularly in connection with excitation transfer,
magnetoreception, vision, and enzyme dynamics
\cite{Cao2020,Marais2018,Alvarez2024}. In photosynthetic
complexes that harvest light, excitation transport is governed by the
interplay between coherent dynamics and environmental fluctuations.
A prominent manifestation of this interplay is quantum transport assisted by the environment
(ENAQT)
\cite{Rebentrost2009,Djordjevic2016,Plenio2008,Caruso2009,Alterman2024},
in which an intermediate level of environmental dephasing can improve
transport by mitigating coherent localization or destructive
interference. ENAQT has consequently become a paradigmatic example of
transport assisted by noise in open quantum systems
\cite{Brixner2005,Ishizaki2009}. Most studies of ENAQT focus on
passive regulation through environmental quantities such as
dephasing strengths, dissipation rates, and noise correlations.
These studies clarify how environmental fluctuations can facilitate
excitation transport. A complementary approach is to actively
manipulate the coherent system dynamics while keeping the
environmental conditions fixed.

Quantum optimal control provides a systematic framework for actively
manipulating quantum dynamics through time-dependent controls
\cite{Brif2010,Glaser2015,Koch2022}. Recent studies have shown, for example, that excitation transport in
open quantum networks can be enhanced by optimizing the on-site
energies, namely, the local excitation energies of the individual
pigment sites \cite{Sgroi2024}. Here, we pursue a complementary strategy by optimizing
the time-dependent inter-site couplings. For a fully connected
network of \(N\) sites, independently controlling every unordered pair of
pigment sites requires \(N(N-1)/2\) control terms, resulting in an
optimization space of high dimension and increasing the complexity of
control implementation. This motivates a central question: can a small,
physically interpretable set of controlled edges retain most of the
transport enhancement achieved under full coupling control? Answering
this question requires a criterion that identifies the edges whose couplings are most relevant to the redistribution of
excitation probability.

To characterize the transport of excitation probability through the open quantum network,
we introduce quantum probability currents, which quantify the instantaneous net transfer of excitation probability
between connected sites \cite{Schumacher2016,Roden2016,Gebauer2004,Salmilehto2012}. Integrating these currents over
a finite evolution time yields the net quantum probability transfer associated with each edge of the open quantum network.

Graph Hodge theory provides a natural geometric framework for analyzing the quantum probability current
network \cite{Jiang2011,Lim2020,Barbarossa2020,Schaub2021}. The integrated quantum probability current can be decomposed into
a gradient component and a cycle component with zero divergence. The gradient component represents net excitation probability
redistribution, whereas the cycle component represents internal circulation that leaves every occupation probability unchanged.
Throughout this work, the occupation probability associated with node \(v\) is denoted by $p_v(t)=\bra v\rho(t)\ket v=\rho_{vv}(t),$
where \(v\in\mathcal{V}\), with \(\mathcal{V}\) denoting the vertex set shown in Figure~\ref{fig:open-network}.


In this work, we combine gradient ascent pulse engineering (GRAPE) \cite{Khaneja2005}
with a Hodge analysis resolved by edge for quantum probability currents integrated over time,
using the change induced by control in the gradient component to rank controllable edges.
We apply this framework to excitation transport in the Fenna--Matthews--Olson (FMO)
complex with seven sites, a protein complex containing pigments in green sulfur bacteria that transfers excitation energy
from the chlorosome antenna toward the reaction center \cite{Muh2007,Blankenship2002}.
The analysis reveals a pronounced reorganization induced by control in the gradient component
of the integrated quantum probability current and uses this redistribution to rank edges for coupling control.
The numerical results distinguish a set of six edges that nearly reproduces the efficiency under full control
from a set of four edges that provides an alternative with lower fluence. Pulse fluence is used here
as a measure of control effort computed after optimization. We use an uncontrolled dephasing scan to examine the
relation between transport efficiency and current, compare the selected set of four edges
with 100 random sets of the same size, and compare the resulting ranking with two reference criteria
on ten random networks with seven sites. Together, these comparisons characterize the tradeoff between efficiency and fluence and examine the
extent to which the ranking remains effective beyond the nominal FMO instance.

The remainder of this paper is organized as follows.
Section II introduces the open quantum transport model,
the coherent control formulation, and the GRAPE
algorithm. Section III develops the criterion based on quantum probability current
for ranking edges under coupling control. Section IV presents the
numerical results for the FMO complex and random network
benchmarks. Section V summarizes the main conclusions and
discusses possible extensions.


\section{Model and dynamics}
\label{sec:model-dynamics}

\subsection{Open Quantum Network Model}

We consider the transport of a single excitation through a
network composed of \(N\) pigment sites. Such a model is commonly used to
describe excitation transfer in photosynthetic protein complexes containing pigments.
The state \(\ket{n}\), with \(n=1,\ldots,N\), represents the
configuration in which the excitation is localized at pigment site \(n\).
Accordingly, the subspace with a single excitation in the pigment network is
\begin{equation*}
\mathcal{H}_{\mathrm{net}}
=
\Span_{\mathbb{C}}
\Set{\ket{1},\ldots,\ket{N}}.
\end{equation*}

\begin{figure}[!htbp]
\centering
\includegraphics{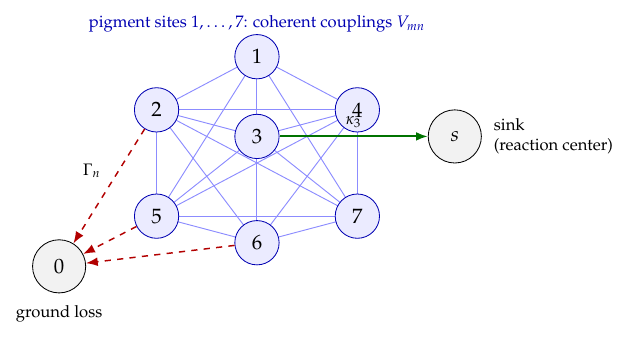}
\caption{Open network model. The numbered circles \(1,\ldots,7\)
denote pigment sites, and the blue lines denote their coherent inter-site
couplings. The circle marked \(0\) is the ground loss state. Dashed red
arrows represent recombination from pigment site \(n\) to \(0\) at rate
\(\Gamma_n\); representative arrows are shown to avoid visual clutter.
The circle marked \(s\) is the sink representing successful excitation
delivery to the reaction center, and the green arrow denotes trapping from
pigment site \(3\) into this sink at rate \(\kappa_3\).}
\label{fig:open-network}
\end{figure}

In addition to the pigment sites, we introduce two auxiliary states to
describe the irreversible outcomes of the transport process.
Figure~\ref{fig:open-network} displays the transport network. The state
\(\ket{0}\) represents excitation recombination, through which the
excitation is lost to the environment without reaching the reaction
center. The state \(\ket{s}\) is an effective sink state representing
successful trapping of the excitation by the reaction center. These
states are bookkeeping states for the two possible irreversible outcomes:
unsuccessful recombination and successful energy delivery, respectively.
The complete Hilbert space therefore has dimension \(N+2\) and is given by
\begin{equation*}
\mathcal{H}
=
\Span_{\mathbb{C}}
\Set{
\ket{0},
\ket{1},\ldots,\ket{N},
\ket{s}
}.
\end{equation*}

\subsubsection{Coherent excitation dynamics}

Within the pigment network, the coherent dynamics is governed by the
tight binding Hamiltonian
\begin{equation*}
\bsH_S
=
\sum_{n=1}^{N}
\epsilon_n \ket{n}\bra{n}
+
\sum_{m<n}
V_{mn}
\left(
\ket{m}\bra{n}
+
\ket{n}\bra{m}
\right).
\end{equation*}
Here, \(\epsilon_n\) denotes the on-site energy of pigment site \(n\),
namely, the excitation energy when the excitation is localized at that
site, whereas \(V_{mn}\) denotes the inter-site coupling strength between
pigment sites \(m\) and \(n\).
The coupling term allows the excitation to move coherently between the
two sites. We assume that \(V_{mn}\) is real, as is customary for the FMO
Hamiltonians considered below.

The ground and sink states are not coherently coupled to the pigment
network. Therefore, in the enlarged Hilbert space, the Hamiltonian takes
the block diagonal form
\begin{equation*}
\bsH
=
0\oplus \bsH_S\oplus 0.
\end{equation*}
The two zero blocks indicate that neither \(\ket{0}\) nor \(\ket{s}\)
participates in the coherent Hamiltonian evolution. These states can be
reached only through the irreversible processes introduced below.


\subsubsection{Environmental and irreversible processes}

A photosynthetic excitation is not isolated from its surrounding
protein and solvent environment. Its dynamics therefore cannot be
described by the Hamiltonian alone. We model the environmental effects
using a Lindblad master equation \cite{Gorini1976,Lindblad1976,Breuer2002},
\begin{equation}
\label{master_equation}
\dot{\rho}(t)
=
-\mathrm{i}\left[\bsH,\rho(t)\right]
+
\cL_{\mathrm{deph}}\!\left[\rho(t)\right]
+
\cL_{\mathrm{rec}}\!\left[\rho(t)\right]
+
\cL_{\mathrm{sink}}\!\left[\rho(t)\right],
\end{equation}
where we set \(\hbar=1\). For any operator \(\bsL\), we define the
Lindblad dissipator as
\begin{equation*}
\mathcal{D}[\bsL]\rho
=
\bsL\rho \bsL^{\dagger}
-
\frac{1}{2}
\Set{\bsL^{\dagger}\bsL,\rho},
\end{equation*}
where
\begin{equation*}
[\bsA,\bsB]=\bsA\bsB-\bsB\bsA,
\qquad
\Set{\bsA,\bsB}=\bsA\bsB+\bsB\bsA
\end{equation*}
denote the commutator and anticommutator, respectively.

The first term on the right side of
Eq.~\eqref{master_equation} describes coherent excitation transfer.
The remaining three terms describe distinct environmental processes.

First, local pure dephasing is modeled as
\begin{equation*}
\mathcal{L}_{\mathrm{deph}}[\rho]
=
\sum_{n=1}^{N}
\mathcal{D}\!\left[L_n^{\mathrm{deph}}\right]\rho,
\qquad
L_n^{\mathrm{deph}}
=
\sqrt{\gamma_n}\ket{n}\bra{n},
\end{equation*}
where \(\gamma_n\) is the dephasing rate at site \(n\)\cite{Haken1973,Rips1993}.

Second, excitation recombination is described by
\begin{equation*}
\mathcal{L}_{\mathrm{rec}}[\rho]
=
\sum_{n=1}^{N}
\mathcal{D}\!\left[L_n^{\mathrm{rec}}\right]\rho,
\qquad
L_n^{\mathrm{rec}}
=
\sqrt{\Gamma_n}\ket{0}\bra{n},
\end{equation*}
where \(\Gamma_n\) is the recombination rate at site \(n\).
It represents an unsuccessful transport event because
the excitation disappears before reaching the reaction center.

Finally, successful trapping by the reaction center is modeled by
\begin{equation*}
\mathcal{L}_{\mathrm{sink}}[\rho]
=
\sum_{n=1}^{N}
\mathcal{D}\!\left[L_n^{\mathrm{sink}}\right]\rho,
\qquad
L_n^{\mathrm{sink}}
=
\sqrt{\kappa_n}\ket{s}\bra{n},
\end{equation*}
where \(\kappa_n\) is the trapping rate from pigment site \(n\) into the
sink.

In the present model, the sink and ground states are absorbing states.
The master equation preserves the total probability in the extended
state space that includes both states,
\begin{equation*}
\Tr{\rho(t)}=1.
\end{equation*}


\subsubsection{Energy transfer efficiency}

The purpose of the transport process is to deliver the excitation to the
reaction center.  Let \(\bsP_s=\ket{s}\bra{s}\) denote the projector onto the
sink state.  We therefore define the transport efficiency at the final time as the
occupation probability of the sink state at the final time \(T\),
\begin{equation}
\label{eff}
\eta(T)
=
\Tr{\bsP_s\rho(T)}
=
\bra{s}\rho(T)\ket{s}
\equiv
\rho_{ss}(T).
\end{equation}
Thus, \(\eta(T)\) is the probability that the excitation has been
successfully trapped by time \(T\).

To obtain an equivalent current representation, we evaluate the evolution equation for the exciton occupation probability in the sink state.
Since the Hamiltonian, dephasing, and recombination
terms do not transfer excitation occupation probability into \(\ket{s}\), only the trapping
dissipator contributes to \(\dot{\rho}_{ss}(t)\). Using
\(L_n^{\mathrm{sink}}=\sqrt{\kappa_n}\ket{s}\bra{n}\), we obtain
\begin{equation*}
\dot{\rho}_{ss}(t)
=
\sum_{n=1}^{N}
\kappa_n\rho_{nn}(t).
\end{equation*}
Here, each term \(\kappa_n\rho_{nn}(t)\) is the instantaneous quantum probability
current from site \(n\) into the sink.

Assuming that the sink is initially unoccupied,
\(\rho_{ss}(0)=0\), integration from \(0\) to \(T\) gives
\begin{align}
\eta(T)
&=
\rho_{ss}(T)-\rho_{ss}(0)
\nonumber\\
&=
\int_{0}^{T}
\dot{\rho}_{ss}(t)\,\dif t
\nonumber\\
&=
\int_{0}^{T}
\sum_{n=1}^{N}
\kappa_n\rho_{nn}(t)\,\dif t.
\label{eff_flux}
\end{align}


\subsection{Controlled Dynamics and GRAPE Optimization}

\begin{figure}[!htbp]
\centering
\includegraphics{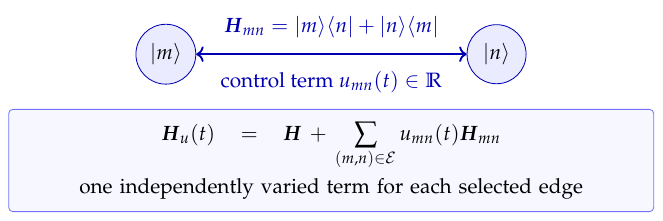}
\caption{Parametrization of coupling control. A selected site pair
\((m,n)\) defines the Hermitian control operator \(\bsH_{mn}\), while
the real control term \(u_{mn}(t)\) modulates that inter-site
coupling.}
\label{fig:control-parameterization}
\end{figure}

We now specialize to time-dependent coupling control, which manipulates
the coherent excitation dynamics while keeping the dephasing,
recombination, and trapping rates fixed.  The controlled Hamiltonian is
\begin{equation}
\label{eq:controlled_hamiltonian}
\bsH_u(t)
=
\bsH+\bsH_c(t).
\end{equation}
Here, \(\bsH\) is the uncontrolled Hamiltonian introduced above and
\(\bsH_c(t)\) is the Hamiltonian for coupling control.  For a controllable
connection between sites \(m\) and \(n\), with \(m<n\), the Hermitian
control operator is
\begin{equation*}
\bsH_{mn}=\ket{m}\bra{n}+\ket{n}\bra{m},\quad m,n=1,2,\cdots,N.
\end{equation*}
Let \(\mathcal E\) denote the set of controllable site pairs and let
\(u_{mn}(t)\) denote the real control term associated with edge
\((m,n)\); it may take either positive or negative values. The
Hamiltonian for coupling control is then
\begin{equation}
\label{eq:coupling_controlled_hamiltonian}
\bsH_c(t)=
\sum_{(m,n)\in\mathcal{E}} u_{mn}(t)\bsH_{mn},
\end{equation}
so every selected connection defines one independent control term.
The operators \(\bsH_{mn}\) act only within the subspace with one excitation
and do not directly couple a pigment site to the ground loss or
sink state. The parametrization of coupling control is summarized in
Figure~\ref{fig:control-parameterization}. If every pair of the \(N\) pigment sites is controllable, then
\begin{equation*}
\mathcal{E}
=
\{(m,n) \mid 1 \leqslant m<n \leqslant N\},
\qquad
M=\lvert\mathcal{E}\rvert
=
\frac{N(N-1)}{2}.
\end{equation*}

Instead of rewriting the complete Lindblad master equation given in the
preceding subsection, we collect its coherent and dissipative contributions
into a Liouvillian superoperator that depends on the controls. The controlled dynamics
can then be written as
\begin{equation}
\label{eq:controlled_liouvillian}
\dot{\rho}(t)
=
\mathcal{L}_{u(t)}\rho(t),
\end{equation}
where
\begin{equation*}
\mathcal{L}_{u(t)}
=
\mathcal{L}_{0}
+
\sum_{(m,n)\in\mathcal E}u_{mn}(t)\mathcal{L}_{mn},
\qquad
\mathcal{L}_{mn}\rho
=
-\mathrm{i}[\bsH_{mn},\rho].
\end{equation*}
Here, \(\mathcal{L}_{0}\) contains the uncontrolled Hamiltonian evolution
and all fixed dissipative terms. Thus, the controls modify only the
coherent part of the dynamics, whereas the dephasing, recombination, and
trapping processes remain unchanged during the optimization.

Our objective is to determine the control terms
\(\{u_{mn}(t)\}_{(m,n)\in\mathcal E}\) that maximize the
transfer efficiency \(\eta(T)\) at the final time in Eq.~\eqref{eff}, subject to the
controlled dynamics in Eq.~\eqref{eq:controlled_liouvillian}.  The control
terms affect \(\eta(T)\) indirectly by changing the evolution of
\(\rho(t)\).

\subsubsection{Augmented functional}

To optimize the control terms, the master equation must be incorporated
as a constraint. For this purpose, we introduce a time-dependent operator
Lagrange multiplier \(\lambda(t)\). The difference
$\dot{\rho}(t) - \mathcal{L}_{u(t)}\rho(t)$
measures how far a trial trajectory \(\rho(t)\) deviates from the master
equation. For a physical trajectory, this difference is identically zero.
Because this constraint is an operator equation, we multiply it by
\(\lambda^\dagger(t)\) and take the trace to obtain a scalar constraint.
The resulting augmented functional is
\begin{equation}
\label{Lagrange_functional}
\eta'[\rho,\lambda,u]
=
\eta(T)
+
\int_{0}^{T}
\Tr{
\lambda^\dagger(t)
\left[\dot{\rho}(t)-
\cL_{u(t)}\rho(t)
\right]
}
\dif t .
\end{equation}

For every trajectory satisfying
Eq.~\eqref{eq:controlled_liouvillian}, the expression inside the integral
vanishes, and therefore
\begin{equation*}
\eta'=\eta.
\end{equation*}
The advantage of Eq.~\eqref{Lagrange_functional} is that
\(\rho(t)\), \(\lambda(t)\), and the active control terms \(u_{mn}(t)\) can now be varied
independently. Variation with respect to \(\rho(t)\), together with the fixed initial
condition \(\delta\rho(0)=0\), gives
\begin{equation}
\label{eq:adjoint_equation}
\dot{\lambda}(t)
=
-\mathcal{L}_{u(t)}^\dagger\lambda(t),
\qquad
\lambda(T)=-\bsP_s.
\end{equation}
A detailed variation of the augmented functional leading to
Eq.~\eqref{eq:adjoint_equation} and its terminal condition is given in
Appendix~\ref{app:grape}.

The superoperator \(\mathcal{L}_{u(t)}^\dagger\) is the adjoint of
\(\mathcal{L}_{u(t)}\) with respect to the Hilbert--Schmidt inner product.
It is defined by the relation
\begin{equation*}
\Tr{
\bsA^\dagger\mathcal{L}_{u(t)}(\bsB)
}
=
\Tr{
\left[
\mathcal{L}_{u(t)}^\dagger(\bsA)
\right]^\dagger \bsB
}
\end{equation*}
for arbitrary operators \(\bsA\) and \(\bsB\).

\begin{figure}[!htbp]
\centering
\includegraphics{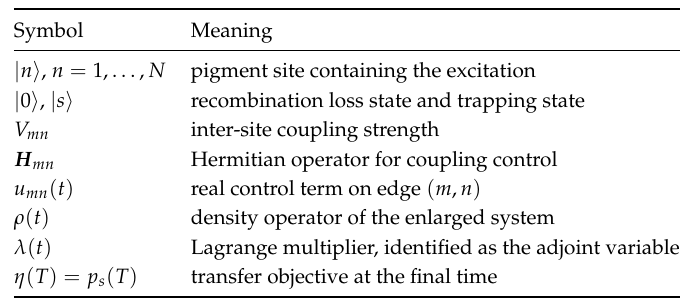}
\caption{Notation used for the open network model and the
optimization of coupling control.}
\label{fig:model-notation}
\end{figure}

Once the stationarity condition has produced
Eq.~\eqref{eq:adjoint_equation}, the auxiliary variable \(\lambda(t)\) is
identified as the adjoint variable, or costate.
The physical state \(\rho(t)\) is propagated forward from its prescribed
initial condition, whereas \(\lambda(t)\) is propagated backward from the
terminal operator \(-\bsP_s\). The two propagations play complementary
roles: \(\rho(t)\) describes where the excitation is at time \(t\), while
\(\lambda(t)\) quantifies how a small change of the state at that time
would affect the final value \(\eta(T)\). Figure~\ref{fig:model-notation} collects the principal symbols after
their introduction in the model and optimization equations.


\subsubsection{Time Discretization and GRAPE Iteration}

Each active controlled edge is labeled by a controllable site pair
\((m,n)\in\mathcal E\). For the numerical implementation, the total
evolution time \(T\) is divided into \(K\) equal intervals,
\begin{equation*}
0=t_0<t_1<\cdots<t_K=T,
\qquad
\Delta t=\frac{T}{K}.
\end{equation*}
Each control term is taken to be constant within one time interval,
\begin{equation*}
u_{mn}(t)=u_{mn,k},
\qquad
t\in[t_k,t_{k+1}),
\end{equation*}
where \((m,n)\in\mathcal E\) labels the controlled site pair and
\(k=0,\ldots,K-1\) labels the time interval. Therefore, all sampled
values of the control terms can be stored in an \(M\times K\) control matrix,
\begin{equation*}
\bsU=
\left(u_{mn,k}\right)\in\mathbb{R}^{M\times K}.
\end{equation*}
Only control terms on the selected controlled edges are optimized; the
control terms on all other edges remain zero throughout the iteration.

Let \(r\) denote the GRAPE iteration number. For a given control matrix
\(\bsU^{(r)}\), the Liouvillian on time interval \(k\) is
\begin{equation*}
\mathcal{L}_{k}^{(r)}
=
\mathcal{L}_{0}
+
\sum_{(m,n)\in\mathcal E}
u_{mn,k}^{(r)}\mathcal{L}_{mn},
\qquad
\mathcal{L}_{mn}\rho
=
-\mathrm{i}[\bsH_{mn},\rho].
\end{equation*}

\paragraph{Step 1: Forward propagation.}

Starting from the prescribed initial state
\(\rho_0=\rho(0)\), the density operator is propagated forward through all
time intervals according to
\begin{equation*}
\rho_{k+1}^{(r)}
=
\exp\left(
\mathcal{L}_{k}^{(r)}\Delta t
\right)
\rho_{k}^{(r)},
\qquad
k=0,\ldots,K-1.
\end{equation*}
This forward propagation produces the complete state trajectory
\begin{equation*}
\left\{
\rho_0^{(r)},\rho_1^{(r)},\ldots,\rho_K^{(r)}
\right\}.
\end{equation*}
The transfer efficiency associated with the current control terms is then
evaluated from the final state \(\rho_K^{(r)}\) using Eq.~\eqref{eff}.

\paragraph{Step 2: Backward propagation.}

The adjoint variable is initialized at the final time using the terminal
condition
\begin{equation*}
\lambda_K^{(r)}=-\bsP_s.
\end{equation*}
It is then propagated backward from \(t_K=T\) to \(t_0=0\) according to
\begin{equation*}
\lambda_k^{(r)}
=
\exp\left[
\left(\mathcal{L}_{k}^{(r)}\right)^\dagger
\Delta t
\right]
\lambda_{k+1}^{(r)},
\qquad
k=K-1,\ldots,0.
\end{equation*}
The backward propagation produces the adjoint trajectory
\begin{equation*}
\left\{
\lambda_0^{(r)},\lambda_1^{(r)},\ldots,\lambda_K^{(r)}
\right\}.
\end{equation*}

The forward state \(\rho_k^{(r)}\) contains information about how the
initial excitation reaches time \(t_k\), whereas
\(\lambda_{k+1}^{(r)}\) contains information about how a change introduced
during interval \(k\) affects the final value \(\eta(T)\). Combining
these two quantities therefore makes it possible to determine the effect
of every control term on the final transfer efficiency.

\paragraph{Step 3: Gradient evaluation.}

The numerical derivative is evaluated with a midpoint discrete GRAPE
rule.  Let \(\rho_{k+1/2}^{(r)}\) and
\(\lambda_{k+1/2}^{(r)}\) denote the forward and adjoint variables
propagated by \(\Delta t/2\) within interval \(k\), over which the Liouvillian
is constant. Then
\begin{equation}
\label{eq:grape_gradient_general}
\frac{\partial\eta^{(r)}}{\partial u_{mn,k}^{(r)}}
=
-\Delta t\,
\operatorname{Re}
\left\langle
\lambda_{k+1/2}^{(r)},
\mathcal{L}_{mn}\rho_{k+1/2}^{(r)}
\right\rangle_{\mathrm{HS}}.
\end{equation}
The variation with respect to \(u_{mn,k}^{(r)}\) and the midpoint
discretization leading to Eq.~\eqref{eq:grape_gradient_general} are
derived in Appendix~\ref{app:grape}; see
Eq.~\eqref{eq:appendix-midpoint-gradient}.
For the controlled edge \((m,n)\),
\begin{equation*}
\mathcal{L}_{mn}\rho
=
-\mathrm{i}[\bsH_{mn},\rho],
\end{equation*}
Eq.~\eqref{eq:grape_gradient_general} becomes
\begin{equation*}
g_{mn,k}^{(r)}
\equiv
\frac{\partial\eta^{(r)}}{\partial u_{mn,k}^{(r)}}
=
-\Delta t\,
\operatorname{Re}
\left\langle
\lambda_{k+1/2}^{(r)},
-\mathrm{i}[\bsH_{mn},\rho_{k+1/2}^{(r)}]
\right\rangle_{\mathrm{HS}}.
\end{equation*}
The collection of all \(g_{mn,k}^{(r)}\) forms an
\(M\times K\) gradient matrix with the same dimensions as the control
matrix \(\bsU^{(r)}\).

A positive value of \(g_{mn,k}^{(r)}\) means that increasing
\(u_{mn,k}^{(r)}\) will increase the final value \(\eta(T)\) to first order.
A negative value means that the control term should instead be
decreased.

\paragraph{Step 4: Control update.}

Because the objective is to maximize the transfer efficiency, the first
stage uses a projected ascent direction normalized by its maximum magnitude,
\begin{equation*}
d_{mn,k}^{(r)}=
 \frac{g_{mn,k}^{(r)}}
 {\max_{(p,q)\in\mathcal A,\,\ell}|g_{pq,\ell}^{(r)}|},
\qquad (m,n)\in\mathcal A,
\end{equation*}
where \(\mathcal A\subseteq\mathcal E\) is the set of actively
optimized edges; all control terms \(u_{mn}(t)\) with \((m,n)\notin \mathcal A\) are fixed at zero.  A trial
update is
\begin{equation*}
\widetilde{u}_{mn,k}^{(r+1)}
=
u_{mn,k}^{(r)}+\alpha_r d_{mn,k}^{(r)},
\end{equation*}
where \(\alpha_r>0\) is the step size. It is adjusted as needed so that the
projected update increases \(\eta(T)\).

To keep the control terms within the prescribed range
\(\lvert u_{mn,k}\rvert\leq u_{\max}\), the updated controls are projected
onto the admissible interval,
\begin{equation*}
u_{mn,k}^{(r+1)}
=
\operatorname{clip}
\left(
\widetilde{u}_{mn,k}^{(r+1)},
-u_{\max},
u_{\max}
\right)
=
\begin{cases}
-u_{\max},
& \widetilde{u}_{mn,k}^{(r+1)}<-u_{\max},\\
\widetilde{u}_{mn,k}^{(r+1)},
& -u_{\max}\leq\widetilde{u}_{mn,k}^{(r+1)}\leq u_{\max},\\
u_{\max},
& \widetilde{u}_{mn,k}^{(r+1)}>u_{\max}.
\end{cases}
\end{equation*}

\paragraph{Step 5: Iteration.}

The forward propagation, backward propagation, gradient evaluation, and
control update are repeated until the efficiency stops improving or the
iteration limit is reached.

\paragraph{Stage 2: Bound constrained refinement.}

The first GRAPE stage provides a feasible control pulse that already improves
the transfer efficiency. However, because GRAPE updates the pulse mainly
according to the local control gradient, its improvement may become slow when the
control is close to a local optimum. We therefore use L-BFGS-B, a limited memory form of the
Broyden--Fletcher--Goldfarb--Shanno algorithm designed for optimization
with bound constraints, to further refine the GRAPE result.

L-BFGS-B is a local optimization method based on gradients and designed for problems
with upper and lower bounds. Starting from the pulse obtained by GRAPE, it
uses the efficiency, the gradient, and the changes observed during recent
iterations to determine a more effective update direction. In this way, it
can adjust the discretized control terms of all active controlled edges together
and seek a further increase in \(\eta(T)\).

\begin{figure}[!htbp]
\centering
\includegraphics{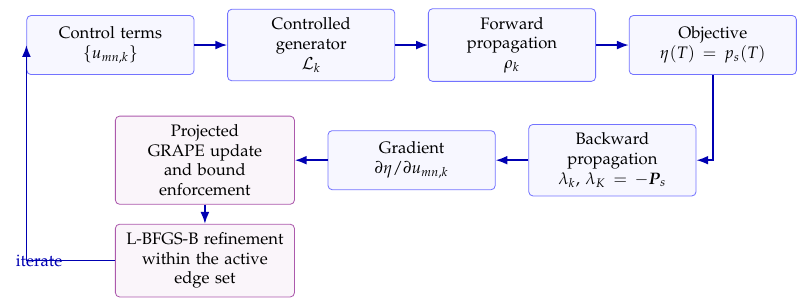}
\caption{Optimization workflow with two stages. The forward state and backward
adjoint trajectories determine the GRAPE gradient; projected GRAPE
updates are followed by L-BFGS-B refinement with bound constraints.}
\label{fig:optimization-workflow}
\end{figure}

During the refinement, every active control term satisfies
\[
\lvert u_{mn,k}\rvert\leqslant u_{\max},
\]
while all inactive edges remain fixed at zero. Therefore, L-BFGS-B does
not select new controlled edges; it only improves the pulse within the
set of controlled edges chosen before the optimization. This stage is a local
refinement, so its result depends on the GRAPE pulse used as the initial
point and does not guarantee a global optimum. The complete optimization procedure with two stages is summarized in
Figure~\ref{fig:optimization-workflow}.

\section{Quantum Probability Current and Hodge Decomposition}
\label{sec:current_hodge}

To understand how the GRAPE control modifies the excitation transport pathways,
we represent the dynamics of the open system in terms of quantum probability currents defined
on the edges of the quantum network \cite{Schumacher2016,Roden2016,Gebauer2004,Salmilehto2012}.
We then integrate these currents over the complete evolution time and apply a graph Hodge
decomposition to separate the gradient and cycle components of the resulting quantum probability current.

\subsection{Quantum Probability Current}

For the controlled Hamiltonian \(\bsH_u(t)\), the coherent quantum probability
current from pigment site \(m\) to pigment site \(n\) is defined as
\begin{equation}
\label{eq:coherent_current}
J_{n\leftarrow m}^{(H)}(t)
=
2\operatorname{Im}
\left[
H_{nm}^{u}(t)\rho_{mn}(t)
\right]
,
\end{equation}
where \(H_{nm}^{u}(t)=\bra n\bsH_u(t)\ket m\), and
\(\operatorname{Im}[\,\cdot\,]\) denotes the imaginary part.
The notation \(J_{n\leftarrow m}^{(H)}(t)\) denotes the instantaneous quantum probability current
from site \(m\) to site \(n\), quantifying the rate of transfer of excitation probability between the two sites at time \(t\).

Because both \(\bsH_u(t)\) and \(\rho(t)\) are Hermitian, their matrix elements satisfy
\begin{equation*}
H_{mn}^{u}(t)
=
\left[H_{nm}^{u}(t)\right]^{*},
\qquad
\rho_{nm}(t)
=
\rho_{mn}^{*}(t),
\end{equation*}
where \({}^{*}\) denotes complex conjugation. Consequently, the
coherent quantum probability current is real and antisymmetric:
\begin{equation*}
J_{n\leftarrow m}^{(H)}(t)
=
-J_{m\leftarrow n}^{(H)}(t).
\end{equation*}

Thus, the two directional notations describe the same signed edge current
rather than two independent physical currents.

In the present numerical implementation, the GRAPE control term enters
the time-dependent inter-site coupling. For a controlled edge connecting sites
\(m\) and \(n\),
\begin{equation*}
H_{nm}^{u}(t)
=
H_{nm}+u_{mn}(t),
\end{equation*}
and Eq.~\eqref{eq:coherent_current} reduces to
\begin{equation}
\label{eq:controlled_coherent_current}
J_{n\leftarrow m}^{(H)}(t)
=
2\left[
H_{nm}+u_{mn}(t)
\right]\operatorname{Im}\rho_{mn}(t).
\end{equation}
Equation
\eqref{eq:controlled_coherent_current} shows that coupling control changes
the current in two ways: it directly changes the controlled coupling term and
indirectly changes the coherence \(\rho_{mn}(t)\) through the controlled
dynamics.

The local pure dephasing operators do not directly transfer excitation probability
between different sites. Therefore,
\begin{equation*}
J_{n\leftarrow m}^{(\mathrm{deph})}(t)
=
0.
\end{equation*}
Nevertheless, dephasing modifies the off diagonal elements of the density matrix
and therefore affects the coherent quantum probability current in
Eq.~\eqref{eq:coherent_current} indirectly.

\begin{figure}[!htbp]
\centering
\includegraphics{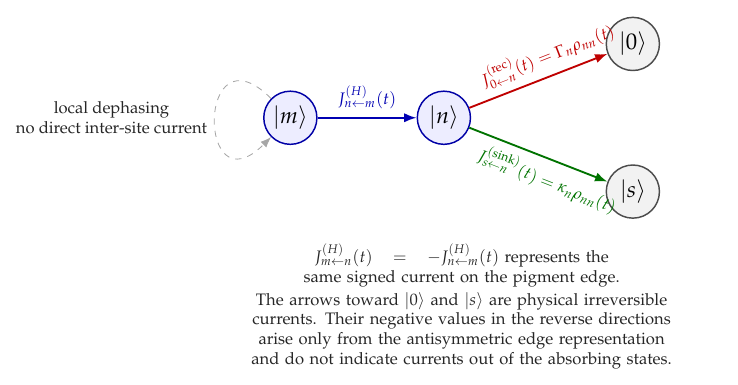}
\caption{Quantum probability current convention for a representative
edge between two pigment sites and its irreversible connections. The blue arrow fixes
the positive orientation of the coherent edge current; a negative value means
that the net current is opposite to this orientation. Recombination and
trapping lead from pigment site $n$ to the absorbing states $\ket{0}$ and
$\ket{s}$, respectively, whereas local dephasing affects the coherent
current only indirectly through the density matrix.}
\label{fig:quantum-current-convention}
\end{figure}

The irreversible current from pigment site \(n\) into the sink is
\begin{equation*}
J_{s\leftarrow n}^{(\mathrm{sink})}(t)
=
\kappa_n\rho_{nn}(t),\quad n=1,\ldots,N,
\end{equation*}
whereas the recombination current from site \(n\) into the ground loss
state is
\begin{equation*}
J_{0\leftarrow n}^{(\mathrm{rec})}(t)
=
\Gamma_n\rho_{nn}(t),\quad n=1,\ldots,N.
\end{equation*}
The values for the reverse directions are introduced only through antisymmetry:
\begin{align*}
J_{n\leftarrow s}^{(\mathrm{sink})}(t)
&=
-J_{s\leftarrow n}^{(\mathrm{sink})}(t),
\\
J_{n\leftarrow0}^{(\mathrm{rec})}(t)
&=
-J_{0\leftarrow n}^{(\mathrm{rec})}(t),\quad n=1,\ldots,N.
\end{align*}
These signed values for the reverse directions do not describe additional
physical currents out of the absorbing states; they are only the
antisymmetric representation of the same oriented sink and recombination
edges. Figure~\ref{fig:quantum-current-convention} summarizes the sign
convention and distinguishes the coherent inter-site current from the
irreversible currents into the two absorbing states.

\subsection{\texorpdfstring{Quantum Probability Current Integrated over Time}{Quantum Probability Current Integrated over Time}}

To characterize the net transport occurring during the complete time
interval \([0,T]\), we define the current integrated over time from node \(m\)
to node \(n\) as
\begin{equation}
\label{eq:integrated_current_matrix}
F_{nm}
=
\int_0^T
J_{n\leftarrow m}(t)\dif t, \quad
m,n \in \{0, 1,\ldots,N, s\}, m \neq n.
\end{equation}

Collecting all integrated currents gives a real antisymmetric current matrix
\begin{equation*}
\bsF
=
(F_{nm})
\in
\mathbb{R}^{(N+2)\times(N+2)}.
\end{equation*}

Using the same time discretization as in the GRAPE propagation,
Eq.~\eqref{eq:integrated_current_matrix} is evaluated numerically by
Gauss--Legendre quadrature with three points on every time interval:
\begin{equation*}
F_{nm}
\approx
\sum_{k=0}^{K-1}
\frac{\Delta t}{2}
\sum_{q=1}^{3}w_q
J_{n\leftarrow m}\!\left(
t_k+\frac{\Delta t}{2}(1+x_q)
\right),
\end{equation*}
where \((x_1,x_2,x_3)=(-\sqrt{3/5},0,\sqrt{3/5})\) and
\((w_1,w_2,w_3)=(5/9,8/9,5/9)\).

The quantity \(F_{nm}\) is a signed quantum probability current integrated over time.
If the current changes direction during the evolution, the forward and backward contributions partially
cancel in the time integral. Thus, \(\bsF\) characterizes the net current accumulated over \([0,T]\),
rather than the total current magnitude accumulated during the evolution.

For the sink edge,
\begin{equation*}
F_{sn}
=
\int_0^T
\kappa_n\rho_{nn}(t)\dif t.
\end{equation*}
Consequently, for an initially unoccupied sink,
\begin{equation*}
\sum_{n=1}^{N}F_{sn}
=
\rho_{ss}(T)
=
\eta(T).
\end{equation*}
Thus, the transfer efficiency is equal to the total integrated quantum probability
current entering the sink.

\subsection{Edge Current Vector and Incidence Matrix}
\label{subsec:incidence_matrix}
Because \(F_{nm}\) and \(F_{mn}\) describe the same edge current with
opposite signs, only one of them is independent. We therefore choose one
fixed reference orientation for each edge of the transport graph. For an
edge with reference direction
\begin{equation*}
e=(m\to n),
\end{equation*}
the corresponding edge current value is defined as
\begin{equation*}
f_e
=
F_{nm}
=
\int_0^T
J_{n\leftarrow m}(t)\dif t.
\end{equation*}
A positive value \(f_e>0\) means that the net integrated current follows the
chosen direction \(m\to n\), whereas \(f_e<0\) means that the actual net
current is in the opposite direction.

Collecting the integrated currents associated with all edges gives the
edge current vector
\begin{equation*}
\bsf
=
\left(
f_1,f_2,\ldots,f_E
\right)^T
\in\mathbb{R}^{E},
\end{equation*}
where $E$ denotes the number of edges. The same edge ordering and reference orientations are used for both the
uncontrolled and controlled dynamics. The transport graph contains the
connections between pigment sites, the trapping edges connecting pigment sites
to the sink, and the recombination edges connecting pigment sites to the
ground state. It has $V=N+2$ nodes: the \(N\) pigment sites, the ground state, and the sink state. Its
incidence matrix is defined as $\bsB\in\mathbb{R}^{V\times E}$. Each row of \(\bsB\) corresponds to one node, and each column corresponds
to one oriented edge. Consider edge \(e\) with reference direction
\(e=(m\to n)\). The elements in the corresponding column of \(\bsB\) are
defined by
\begin{equation*}
B_{ve}
=
\begin{cases}
-1, & v=m,\\[2pt]
+1, & v=n,\\[2pt]
0,  & v\neq m,n.
\end{cases}
\end{equation*}

The incidence matrix contains only information about the topology and
the selected edge orientations. It does not contain the magnitude of the
quantum probability current. The current magnitudes are stored separately in the vector
\(\bsf\). Multiplying \(\bsB\) by \(\bsf\) distributes the edge currents
among the nodes:
\begin{equation*}
\bsB\bsf
\in\mathbb{R}^{V}.
\end{equation*}
The component corresponding to node \(v\) is
\begin{equation*}
(\bsB\bsf)_v
=
\sum_{e=1}^{E}B_{ve}f_e.
\end{equation*}
Because an entering edge contributes \(+f_e\) and a leaving edge
contributes \(-f_e\), this component is the total integrated inflow into
node \(v\) minus the total integrated outflow:
\begin{equation*}
(\bsB\bsf)_v
=
\sum_{\substack{e\,\mathrm{entering}\\\mathrm{node}\ v}}f_e
-
\sum_{\substack{e\,\mathrm{leaving}\\\mathrm{node}\ v}}f_e.
\end{equation*}

To relate this net current to the dynamics of the occupation probabilities, we introduce the
instantaneous occupation probability vector
\begin{equation*}
\bsp(t)
=
\left(
p_0(t),p_1(t),\ldots,p_N(t),p_s(t)
\right)^T
\in\mathbb{R}^{V},
\end{equation*}
where the components \(p_v(t)\) were defined in the Introduction. Specifically,
\(p_n(t)\) is the probability that the excitation is located at pigment
site \(n\), \(p_0(t)\) is the probability that it has been lost through
recombination, and \(p_s(t)\) is the probability that it has been trapped
by the sink.

Using the same edge ordering as in \(\bsf\), we define the instantaneous
edge current vector as
\begin{equation*}
\bsj(t)
=
\left(
j_1(t),j_2(t),\ldots,j_E(t)
\right)^T,
\end{equation*}
where, for \(e=(m\to n)\),
\begin{equation*}
j_e(t)
=
J_{n\leftarrow m}(t).
\end{equation*}
For any node \(v\), the continuity equation is expressed as
\begin{align*}
\dot{p}_v(t) = \sum_{m\ne v} J_{v\leftarrow m}(t).
\end{align*}
The continuity equation for the occupation probabilities can therefore be written in matrix form as
\begin{equation}
\label{eq:instantaneous_continuity_matrix}
\dot{\bsp}(t)
=
\bsB\bsj(t).
\end{equation}

Integrating Eq.~\eqref{eq:instantaneous_continuity_matrix} from \(0\) to
\(T\) gives
\begin{align*}
\bsp(T)-\bsp(0)
&=
\int_0^T
\dot{\bsp}(t)\dif t
\nonumber\\
&=
\bsB
\int_0^T
\bsj(t)\dif t
\nonumber\\
&=
\bsB\bsf.
\end{align*}
Therefore,
\begin{equation*}
\bsB\bsf
=
\bsp(T)-\bsp(0)
.
\end{equation*}
This relation shows that the net integrated inflow at each node is exactly
equal to the change in the occupation probability of that node between times \(0\)
and \(T\).

The notation \((\bsB\bsf)_s\) denotes the component of
\(\bsB\bsf\) associated with the sink node. In the present transport
graph, all trapping edges are oriented from the pigment sites into the
sink. Because \(p_s(0)=0\), the sink component of the integrated
continuity equation becomes
\begin{equation*}
(\bsB\bsf)_s=p_s(T)-p_s(0)=p_s(T)=\eta(T).
\end{equation*}

\subsection{Hodge Decomposition}
\label{subsec:hodge_decomposition}

The edge current space admits the orthogonal decomposition
\cite{Jiang2011,Lim2020}
\begin{equation*}
\mathbb{R}^{E}
=
\operatorname{range}(\bsB^T)
\oplus
\ker(\bsB).
\end{equation*}
Here,
\begin{align*}
\ker(\bsB)
&=\left\{\bsx\in\mathbb{R}^{E}:\bsB\bsx=\boldsymbol{0}\right\},\\
\operatorname{range}(\bsB^T)
&=\left\{\bsB^T\phi:\phi\in\mathbb{R}^{V}\right\}
\subseteq\mathbb{R}^{E}.
\end{align*}

Accordingly, the integrated current has the unique decomposition
\begin{equation*}
\bsf
=
\bsf_{\mathrm{grad}}
+
\bsf_{\mathrm{cycle}}
,
\end{equation*}
where \(\bsf_{\mathrm{grad}}\) is the gradient component belonging to
\(\operatorname{range}(\bsB^T)\), and \(\bsf_{\mathrm{cycle}}\) is the
divergence-free cycle component belonging to \(\ker(\bsB)\).
These components satisfy
\begin{equation*}
\bsB\bsf_{\mathrm{grad}}
=
\bsB\bsf
=
\bsp(T)-\bsp(0),
\qquad
\bsB\bsf_{\mathrm{cycle}}
=
0.
\end{equation*}
Hence, the gradient component contains the complete redistribution of the occupation probabilities
at every node, whereas the cycle component represents
circulation with zero divergence. In particular,
\begin{equation*}
\eta(T)
=
\left(
\bsB\bsf_{\mathrm{grad}}
\right)_s,
\end{equation*}
while the cycle component makes no direct contribution to the final sink
value \(\eta(T)(p_s(T))\).
To determine the gradient component, write
\(\bsf_{\mathrm{grad}}=\bsB^T\phi\), where
\(\phi\in\mathbb{R}^{V}\). The
orthogonal projection of \(\bsf\) onto
\(\operatorname{range}(\bsB^T)\) is obtained from
\cite{Jiang2011,Schaub2018}
\begin{equation*}
\min_{\phi}
\left\|
\bsf-\bsB^T\phi
\right\|_2^2.
\end{equation*}
The associated equation is
\begin{equation*}
\bsB\bsB^T\phi
=
\bsB\bsf.
\end{equation*}
Defining the graph Laplacian
\begin{equation*}
\bsL_{\mathcal G}
=
\bsB\bsB^T,
\end{equation*}
a solution with minimum norm is
\begin{equation*}
\phi
=
\bsL_{\mathcal G}^{+}\bsB\bsf,
\end{equation*}
where the superscript \(+\) denotes the Moore--Penrose pseudoinverse.

The gradient component is therefore
\begin{equation*}
\bsf_{\mathrm{grad}}
=
\bsB^T
\left(
\bsB\bsB^T
\right)^+
\bsB\bsf,
\end{equation*}
and the cycle component is
\begin{equation*}
\bsf_{\mathrm{cycle}}
=
\left[
\bsI_E
-
\bsB^T
\left(
\bsB\bsB^T
\right)^+
\bsB
\right]\bsf.
\end{equation*}

\subsection{Current Measures}
\label{subsec:current_measures}

For comparison with the numerical current matrices, we map the edge
components back to the antisymmetric matrices
\(\bsF_{\mathrm{grad}}\) and \(\bsF_{\mathrm{cycle}}\) using the same
edge orientations as for \(\bsF\). Their magnitudes are defined by
\begin{align*}
G
&=
\left\|
\bsF_{\mathrm{grad}}
\right\|_{\mathrm F},
&
C
&=
\left\|
\bsF_{\mathrm{cycle}}
\right\|_{\mathrm F},
\end{align*}
where
\begin{equation*}
\|\bsA\|_{\mathrm F}
=
\left(
\sum_{i,j}|A_{ij}|^2
\right)^{1/2}.
\end{equation*}
Because every physical edge appears twice in an antisymmetric matrix,
\begin{equation*}
G
=
\sqrt{2}\,
\|\bsf_{\mathrm{grad}}\|_2,
\qquad
C
=
\sqrt{2}\,
\|\bsf_{\mathrm{cycle}}\|_2.
\end{equation*}

The changes induced by control are quantified as
\begin{equation*}
\Delta G
=
G^{(\mathrm c)}-G^{(0)},
\qquad
\Delta C
=
C^{(\mathrm c)}-C^{(0)}
{,}
\end{equation*}
where the superscripts \((0)\) and \((\mathrm c)\) denote the uncontrolled
and controlled dynamics.

The quantities \(\Delta G\) and \(\Delta C\) measure the changes induced by control in the magnitudes
of the gradient and cycle components of the integrated quantum probability current, respectively.
Although \(G\) is not identical to the transfer efficiency, because the gradient component also accounts
for net probability redistribution toward the ground loss state and the remaining pigment sites, the numerical
control cases considered below yield \(\Delta G>0\) together with an enhanced value of \(\eta(T)\). The increase in transfer
efficiency is therefore accompanied by a systematic strengthening of the gradient component of the integrated
quantum probability current. In contrast, \(\Delta C\) has no universal sign, indicating that internal circulation with zero divergence
internal circulation may be either enhanced or suppressed depending on the optimized transport dynamics.
Within these cases, the positivity of \(\Delta G\) provides a common geometric signature of transport enhancement induced by control.

\section{Numerical simulations}
\label{sec:numerical}

\subsection{FMO model and numerical protocol}
\label{subsec:numerical-protocol}

We consider excitation transport in one monomer of the
Fenna--Matthews--Olson (FMO) complex using the seven-site
Adolphs--Renger Hamiltonian \cite{Adolphs2006,Engel2007,Muh2007}.
With the energy of site 3 set to zero, the Hamiltonian in
\(\mathrm{cm}^{-1}\) is
\begin{equation*}
\bsH_S(\mathrm{cm}^{-1})=
\begin{pmatrix}
215 & -104.1 & 5.1 & -4.3 & 4.7 & -15.1 & -7.8\\
-104.1 & 220 & 32.6 & 7.1 & 5.4 & 8.3 & 0.8\\
5.1 & 32.6 & 0 & -46.8 & 1.0 & -8.1 & 5.1\\
-4.3 & 7.1 & -46.8 & 125 & -70.7 & -14.7 & -61.5\\
4.7 & 5.4 & 1.0 & -70.7 & 450 & 89.7 & -2.5\\
-15.1 & 8.3 & -8.1 & -14.7 & 89.7 & 330 & 32.7\\
-7.8 & 0.8 & 5.1 & -61.5 & -2.5 & 32.7 & 280
\end{pmatrix}.
\end{equation*}
The site energies and intersite couplings are taken from the
widely used Adolphs--Renger parametrization of the FMO complex
\cite{Adolphs2006}. The conversion
\(1~\mathrm{cm}^{-1}=0.188365~\mathrm{ps}^{-1}\)
expresses the Hamiltonian in angular-frequency units with \(\hbar=1\).

The Hilbert space also contains the ground-loss state \(\ket{0}\) and
the irreversible sink state \(\ket{s}\). The initial state is
\(\rho(0)=\ket{1}\bra{1}\), and site 3 is coupled to the sink.
Following commonly used FMO transport parametrizations
\cite{Baker2015}, we take the excitation-trapping rate and
recombination rate as
\(\kappa_3=1~\mathrm{ps}^{-1}\) and
\(\Gamma_n=10^{-3}~\mathrm{ps}^{-1}\), respectively.
Unless specified otherwise, we use a fixed dephasing rate
\(\gamma=1~\mathrm{cm}^{-1}\), so that
\begin{equation*}
\gamma=1~\mathrm{cm}^{-1}=0.188365~\mathrm{ps}^{-1},\qquad
\Gamma_n=10^{-3}~\mathrm{ps}^{-1},\qquad
\kappa_3=1~\mathrm{ps}^{-1},
\end{equation*}
with \(\kappa_n=0\) for \(n\ne3\).

The objective is
\begin{equation*}
\eta(T)=p_s(T)=\rho_{ss}(T)
=\int_0^T\kappa_3\rho_{33}(t),\mathrm{d}t.
\end{equation*}

All control calculations use \(T=10~\mathrm{ps}\) and \(K=1000\)
piecewise constant intervals, giving \(\Delta t=0.01~\mathrm{ps}\).
The candidate controllable edges are
\(\mathcal E=\{(m,n):1\le m<n\le7\}\), so full control contains 21
independent terms for coupling control.  Following the notation of
Sec.~\ref{sec:model-dynamics}, the controlled Hamiltonian is
\begin{equation*}
\bsH_u(t)=\bsH_S+\sum_{(m,n)\in\mathcal E}u_{mn}(t)\bsH_{mn},
\qquad
\bsH_{mn}=\ket{m}\bra{n}+\ket{n}\bra{m}.
\end{equation*}
Each real control term obeys
\begin{equation*}
|u_{mn,\ell}|\leqslant18.8365~\mathrm{ps}^{-1}
=100~\mathrm{cm}^{-1},\qquad \ell=0,\ldots,K-1.
\end{equation*}
For an active set \(\mathcal A\subseteq\mathcal E\), control terms outside
\(\mathcal A\) are fixed to zero, while the physical couplings in
\(\bsH_S\) remain unchanged.  The number of control degrees of freedom
below refers to the number of independently driven edges,
\(k=|\mathcal A|\); the corresponding discretized optimization has
\(kK\) variables for the control terms.

The control fluence is evaluated afterwards as
\begin{equation*}
\mathcal F[u]=\sum_{(m,n)\in\mathcal A}
\int_0^T u_{mn}^2(t)\,\mathrm{d}t,
\end{equation*}
with units \(\mathrm{ps}^{-1}\). We use this quantity as a measure of control effort computed after optimization.
Because it is not included in the optimization
objective, the reported values are the observed fluences of the selected
pulses that maximize efficiency rather than minima of a fluence cost function.

To rank the controllable edges, we define the normalized score
\begin{equation}
D_{mn}=
\frac{\left|[\bsF_{\mathrm{grad}}^{(21,0)}
-\bsF_{\mathrm{grad}}^{(0)}]_{mn}\right|}
{\max_{(p,q)\in\mathcal E}
\left|[\bsF_{\mathrm{grad}}^{(21,0)}
-\bsF_{\mathrm{grad}}^{(0)}]_{pq}\right|},
\label{eq:edge-ranking}
\end{equation}
where \(\bsF_{\mathrm{grad}}^{(0)}\) and
\(\bsF_{\mathrm{grad}}^{(21,0)}\) denote the gradient components of the
quantum probability current integrated over time for the uncontrolled and
full control dynamics, respectively. A larger \(D_{mn}\) indicates a
stronger change induced by control in the gradient component on edge
\((m,n)\), and the controllable edges are ranked in descending order of
\(D_{mn}\). If \(\pi(r)\in\mathcal E\)
is the edge ranked in position \(r\), we define
\begin{equation*}
\mathcal S_k=\{\pi(r):r=1,\ldots,k\},\qquad
D_{\pi(1)}\geqslant\cdots\geqslant D_{\pi(21)}.
\end{equation*}
We call \(\mathcal S_k\) the Top-\(k\) edge set; it contains the
\(k\) controllable edges with the largest \(D_{mn}\) values. Accordingly,
a Top-\(k\) result is obtained by restricting the optimization to these
\(k\) edges.

For \(k\in\{1,2,3,4,5,6,7,9,11\}\), the tested candidates include
independent optimization from a zero initial pulse, direct truncation of the full pulse,
optimization of that truncated pulse, and continuation. For each
subsequent tested size, continuation starts from the candidate with the highest efficiency
among the three protocols without continuation at the preceding
tested size, with newly activated control terms initialized to zero.
At the first size, the continuation entry simply reuses the best
candidate obtained without continuation. Direct truncation is an evaluation without
further optimization; all optimized candidates use the solver with two stages
described above.  We report
\begin{equation}
\eta_k=\max_{q\in\mathcal P_k}\eta_k^{(q)},
\label{eq:numerical-best-over-protocols}
\end{equation}
where \(\mathcal P_k\) contains the tested candidates at that \(k\).
For the pulse attaining this largest observed efficiency,
\(u_k^{\mathrm{best}}\), its corresponding fluence is
\begin{equation}
\mathcal F_k=\mathcal F[u_k^{\mathrm{best}}]
=\Delta t\sum_{(m,n)\in\mathcal S_k}\sum_{\ell=0}^{K-1}
\left(u_{mn,\ell}^{\mathrm{best}}\right)^2.
\label{eq:numerical-fluence-k}
\end{equation}
Efficiency and fluence always refer to the same pulse.  This selection
does not minimize fluence over the tested initializations.

\subsection{Efficiency saturation and fluence as a control cost across numbers of edges}
\label{subsec:numerical-full-top4}

The uncontrolled efficiency is \(\eta_0=33.54\%\). Further refinement
of the preliminary pulse under full control in the complete control space with 21 edges
gives the numerical benchmark
\begin{equation}
\eta_{21}^{(\mathrm{bench})}=99.49\%,\qquad
\mathcal F_{21}=659.4~\mathrm{ps}^{-1}.
\label{eq:numerical-full-control}
\end{equation}
The gain retention ratio for the reduced control results is defined as
\begin{equation*}
Q_k=
\frac{\eta_k-\eta_0}
{\eta_{21}^{(\mathrm{bench})}-\eta_0}.
\end{equation*}

This ratio measures the fraction of the enhancement induced by control
retained by the reduced set, excluding the efficiency already present
without control.

\begin{table*}[t]
 \centering
 \caption{Best Top-\(k\) results. For each $k$, $\eta_k$ is the
 largest terminal efficiency found over the protocols using a zero initial pulse,
 projection, optimization after projection, and continuation
 [Eq.~\eqref{eq:numerical-best-over-protocols}].
 $Q_k$ is evaluated from this value.  $\mathcal F_k$ [in
 $\mathrm{ps}^{-1}$; Eq.~\eqref{eq:numerical-fluence-k}] is the
 fluence of the corresponding pulse computed after optimization and not included in
 the objective.}
 \label{tab:numerical-topk-comparison}
 \small
 \begin{tabular}{rccc}
  \toprule
  $k$ & $\eta_k~(\%)$ & $Q_k$ & $\mathcal F_k~(\mathrm{ps}^{-1})$ \\
  \midrule
   1 & 94.40 & 0.9228 & 904.8 \\
   2 & 97.19 & 0.9652 & 931.7 \\
   3 & 97.81 & 0.9744 & 582.8 \\
   4 & 97.91 & 0.9760 & 385.4 \\
   5 & 98.08 & 0.9785 & 448.1 \\
   6 & 99.38 & 0.9983 & 623.9 \\
   7 & 99.41 & 0.9988 & 574.2 \\
   9 & 99.42 & 0.9989 & 559.1 \\
  11 & 99.44 & 0.9992 & 616.9 \\
  \bottomrule
 \end{tabular}
\end{table*}

Table~\ref{tab:numerical-topk-comparison} summarizes the best optimized results for different numbers \(k\) of controlled edges.
Efficiency increases from \(94.40\%\) at \(k=1\) to \(98.08\%\) at
\(k=5\), and reaches \(99.38\%\) at \(k=6\). The set of six edges is
\begin{equation*}
\mathcal S_6=\mathcal S_4\cup\{(3,5),(3,4)\},\qquad
\mathcal S_4=\{(3,7),(1,3),(2,3),(3,6)\}.
\end{equation*}
Its efficiency is only \(0.11\) percentage points below the full control
benchmark, and \(Q_6=0.9983\): approximately \(99.83\%\) of the
full control enhancement is retained with only six independently driven
edges. This is a \(71.43\%\) reduction in the number of control
terms relative to the complete space with 21 edges. Its fluence,
\(623.9~\mathrm{ps}^{-1}\), is \(5.39\%\) below the full control value.
Six edges therefore provide a compact choice when the main requirement is
efficiency close to that of full control.

Beyond \(k=6\), the efficiency gains are small. The Top-$7$, Top-$9$,
and Top-$11$ sets add only \(0.03\), \(0.04\), and \(0.06\) percentage
points, respectively, relative to Top-6. The observed efficiencies
therefore enter a regime close to saturation with six edges. This is
an empirical saturation within the tested sequence, rather than proof
that six edges attain the global maximum or are the minimum number
needed for a specified accuracy.

The solution with four edges offers an alternative with lower fluence, achieving \(\eta_4=97.91\%\)
and retaining \(97.60\%\) of the full control enhancement, with \(\mathcal F_4=385.4~\mathrm{ps}^{-1}\),
the lowest fluence among the solutions with high efficiency listed in
Table~\ref{tab:numerical-topk-comparison}. Relative to full control, it
uses \(80.95\%\) fewer independently driven edges and reduces the
fluence by \(41.55\%\). Relative to Top-$6$, it uses two fewer control terms
and \(38.22\%\) less fluence, at an efficiency decrease of \(1.47\)
percentage points. Top-$4$ thus offers an option with lower fluence when this
efficiency difference is acceptable.

Table \ref{tab:numerical-topk-comparison} shows that using fewer controlled edges does not necessarily reduce the pulse fluence.
The number of active edges and the fluence should therefore be treated as separate measures of control cost.

The corresponding quantum probability currents help interpret the
reduced control sets.  Using
\begin{equation*}
\bsF=\bsF_{\mathrm{grad}}+\bsF_{\mathrm{cycle}},\qquad
G=\|\bsF_{\mathrm{grad}}\|_{\mathrm F},\qquad
C=\|\bsF_{\mathrm{cycle}}\|_{\mathrm F},
\end{equation*}
the uncontrolled, Top-$4$, Top-$6$, and full control trajectories give
\begin{align*}
G_0&=0.6808,&G_4&=1.5514,&G_6=1.5724,\qquad
&G_{21}&=1.5739,\\
C_0&=0.9584,&C_4&=1.2553,&C_6=1.1801,\qquad
&C_{21}&=1.1795.
\end{align*}
Both reduced solutions substantially increase the gradient component norm,
with Top-$6$ close to the full control value. The cycle component norm does not
follow the same ordering as efficiency. These results indicate that selection based on \(D\)
identifies couplings that participate most strongly in the reorganization induced by control
of the gradient component of the quantum probability current integrated over time. The
ranking therefore serves as a structural criterion for selecting candidate
control edges, while the quality of each reduced control subspace is ultimately assessed
by the optimized transfer efficiency.

\subsection{\texorpdfstring{Quantum Probability Current Diagnostics and the Random Baseline with Four Edges}{Quantum Probability Current Diagnostics and the Random Baseline with Four Edges}}
\label{subsec:numerical-external-scope}
We use two supplementary diagnostics to assess the physical
interpretation and practical selectivity of the \(D_{mn}\) ranking. First,
an uncontrolled dephasing scan tests the relation between efficiency and
the quantum probability current decomposition. Second, a matched random
comparison with four edges tests whether the ranked Top-$4$ set provides a
favorable tradeoff between efficiency and fluence.

\begin{figure}[!htbp]
 \centering
 \includegraphics[width=0.77\textwidth]{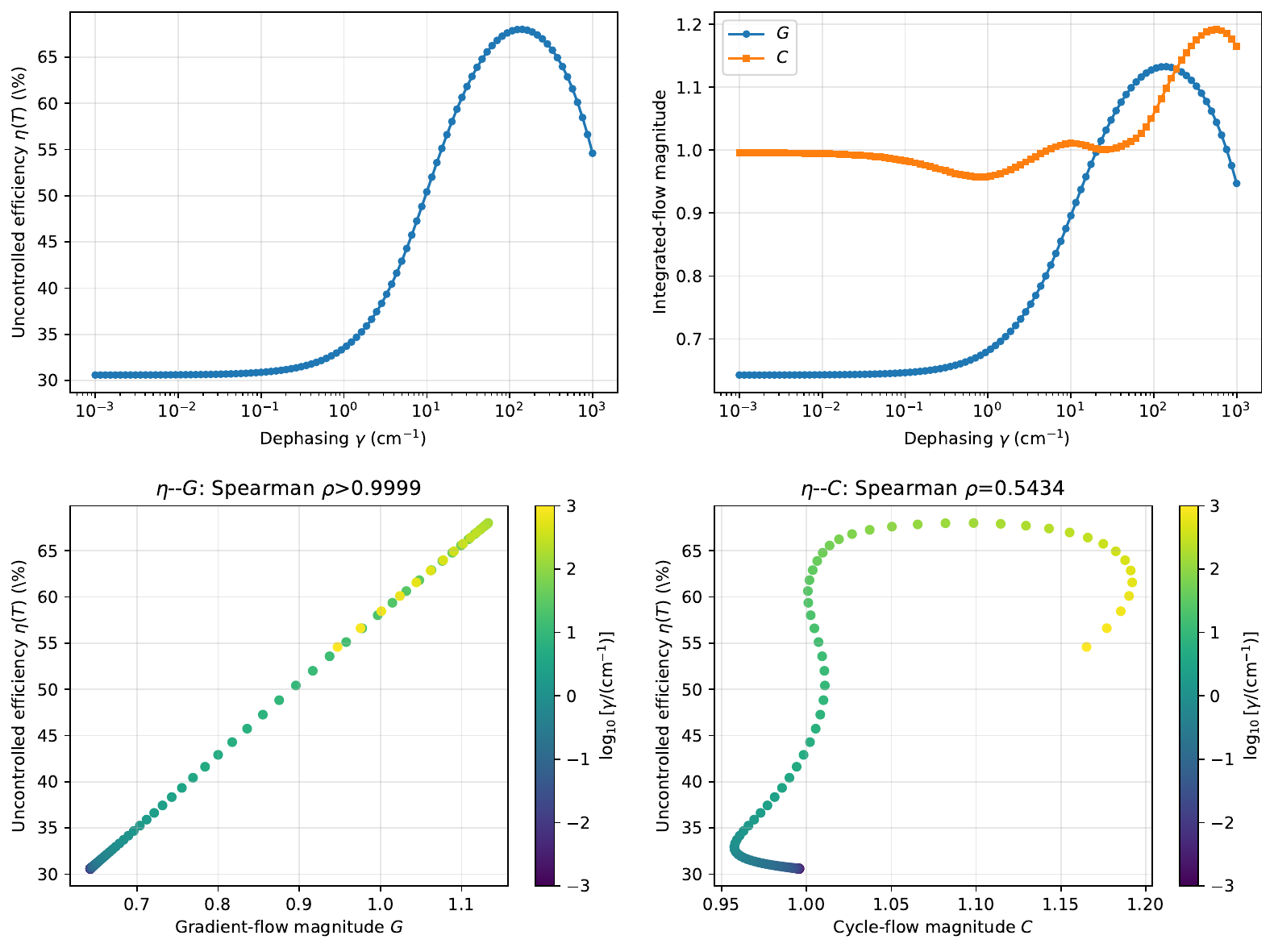}
 \caption{Uncontrolled FMO dephasing scan over 100 logarithmically spaced
 rates.  Top left: terminal sink efficiency versus dephasing rate.
 Top right: norms of the gradient and cycle components of the
 quantum probability current integrated over time versus dephasing rate.
 Bottom: efficiency versus the gradient component norm \(G\) (left)
 and the cycle component norm \(C\) (right); color denotes
 \(\log_{10}\gamma\), with \(\gamma\) expressed in \(\mathrm{cm}^{-1}\).}
 \label{fig:numerical-dephasing-current-scan}
\end{figure}

For the first diagnostic, we examine how the quantum probability
current decomposition varies with transport efficiency in an uncontrolled
dephasing scan. For 100 logarithmically spaced rates
\(\gamma\in[10^{-3},10^3]~\mathrm{cm}^{-1}\), the gradient component
norm closely tracks efficiency, with Spearman rank correlation
\(\rho_{\eta,G}>0.9999\). Here, Spearman's rank correlation coefficient \(\rho\) is a
nonparametric measure of the strength and direction of a monotonic
association between two variables, with \(\rho=1\) and \(\rho=-1\)
denoting perfect increasing and decreasing monotonic relations,
respectively. The cycle component norm has a weaker
monotonic association, \(\rho_{\eta,C}=0.5434\). These correlations
describe a scan over one parameter for the fixed FMO Hamiltonian and do not
establish a universal relation between efficiency and either norm.
Figure~\ref{fig:numerical-dephasing-current-scan} shows the dephasing
dependence and the corresponding scatter plots.

We next compare the Top-$4$ set selected by the \(D_{mn}\) ranking with
100 randomly chosen sets of four edges, all optimized using the same procedure.
The random efficiencies range from \(94.38\%\) to \(99.31\%\), with mean
\(97.16\%\). The ranked Top-$4$ pulse reaches \(97.90\%\),
exceeds 86 of the 100 random results, and lies \(0.74\) percentage points
above their mean. Random pulse
fluences range from \(521.5\) to \(8871.7~\mathrm{ps}^{-1}\), with mean
\(1615.4~\mathrm{ps}^{-1}\). The ranked pulse has fluence
\(435.5~\mathrm{ps}^{-1}\), which is \(73.04\%\) below the random mean and
\(16.48\%\) below the smallest sampled random value.

The corresponding empirical tail fractions with one added count are
\begin{equation*}
p_{\eta}=\frac{1+14}{1+100}=0.1485,\qquad
p_{\mathcal F}=\frac{1+0}{1+100}=0.0099.
\end{equation*}
Here, \(p_{\eta}\) is the fraction of random sets of four edges whose optimized
efficiency is at least as large as that of the set selected by \(D_{mn}\),
whereas \(p_{\mathcal F}\) is the fraction whose fluence is no greater than
that of the set selected by \(D_{mn}\). These results show that the
Top-$4$ set selected by \(D_{mn}\) combines relatively high efficiency with
unusually low fluence within the random ensemble initialized at zero.

\subsection{Extension to random networks}
\label{subsec:numerical-random-network-extension}

To test whether the result from ranking the edges extends beyond the FMO model,
we compared the three ranking criteria for Top-$4$ control across ten
random networks with seven sites. The ensemble results are summarized in
Table~\ref{tab:numerical-random-network-extension}.

\begin{table}[htbp]
\centering
\caption{Top-$4$ comparison averaged over ten random networks with seven sites.}
\label{tab:numerical-random-network-extension}
\small
\begin{tabular}{lccc}
\toprule
Ranking criterion
& Mean \(\eta_4\)
& Mean \(Q_4\)
& Mean \(\mathcal F_4~(\mathrm{ps}^{-1})\)
\\
\midrule
\(D_{mn}\)
& \(95.75\%\)
& \(0.9371\)
& \(2236\)
\\
\(\lvert V_{mn}\rvert\)
& \(95.01\%\)
& \(0.9213\)
& \(3987\)
\\
\(\lvert F_{mn}^{(0)}\rvert\)
& \(94.21\%\)
& \(0.9112\)
& \(3584\)
\\
\bottomrule
\end{tabular}
\end{table}

The selection based on \(D_{mn}\) gives the best aggregate performance,
with the highest mean efficiency and gain retention and the lowest mean
fluence. Although this ensemble advantage does not imply superiority in
every individual network, it shows that \(D_{mn}\) remains an effective
criterion for selecting a compact set of control edges.

\subsection{Numerical consistency and implications}
\label{subsec:numerical-consistency}

For the \(K=1000\) FMO trajectories, the trace and Hermiticity
residuals remain at floating point accuracy. The quantum probability
current balance residual is \(1.98\times10^{-9}\) without control,
\(1.61\times10^{-7}\) for Top-$4$, \(1.38\times10^{-7}\) for Top-$6$,
and \(2.78\times10^{-7}\) for full control. In all four cases, the
divergence residual of the cycle component is below \(1.5\times10^{-15}\), confirming
that this component has zero divergence. These results
show that the trajectory integration and quantum probability current
decomposition are numerically consistent.

The FMO results identify two useful reduced control regimes. Top-$6$
reaches \(99.38\%\) efficiency and retains \(99.83\%\) of the
full control enhancement with only six driven edges. Top-$4$ reaches
\(97.91\%\) efficiency while using \(38.22\%\) less fluence than
Top-$6$ and \(41.55\%\) less than full control. The comparisons with random sets and
random networks further support \(D_{mn}\) as an effective
criterion for selecting edges. Overall, the preferred number of active
edges depends on the desired balance between transfer efficiency and
control fluence.

\section{Conclusion and outlook}
\label{sec:conclusion}

We developed a framework guided by quantum probability current for reducing the number of
independently controlled couplings in open quantum transport. The central
ranking score is the change resolved by edge in the gradient component of the
quantum probability current integrated over time between uncontrolled and
full control dynamics. Graph Hodge decomposition separates this component,
which describes net probability redistribution, from circulation with zero divergence
and thereby provides a physically interpretable basis for
constructing reduced control sets.

For the FMO model with seven sites, control of all 21 edges increases the transfer
efficiency from \(33.54\%\) to \(99.49\%\). The Top-$6$ set reaches
\(99.38\%\) and retains \(99.83\%\) of the full control enhancement with
only six independently driven edges. The Top-$4$ set reaches
\(97.91\%\), retains \(97.60\%\) of the enhancement, and has
\(41.55\%\) lower fluence than full control. Thus, Top-$6$ represents the
regime in which efficiency is close to saturation, whereas Top-$4$ provides the regime
with lower fluence. The close association between efficiency and the
gradient component norm in the uncontrolled dephasing scan is consistent
with the interpretation based on quantum probability current for this FMO
setting, without implying a general causal relation.

The random baseline for the fixed FMO model clarifies the scope of the
result with four edges. Fourteen of 100 random sets of four edges attain higher
efficiency than the set ranked by \(D_{mn}\), but none has lower fluence; the
ranking therefore identifies a favorable tradeoff between efficiency and fluence
rather than guaranteeing the maximum efficiency. Across ten random
networks with seven sites, \(D_{mn}\)
gives the highest mean efficiency and gain retention and the lowest mean
fluence among the three tested ranking criteria. These ensemble results
provide initial evidence that the selection rule remains useful beyond the
nominal FMO instance, while allowing the preferred size of the control set to depend
on the network and the desired balance between performance and fluence.

A larger ensemble of random networks, together with searches from multiple starting points or global
searches, would quantify the robustness of these comparisons against network
variation and local optimization basins.  Future calculations should also
incorporate smoothness, bandwidth, and fluence constraints directly into the
optimization, and extend the Markovian model to static disorder, structured
environments, and dynamics beyond the Markovian approximation. These steps will clarify the
conditions under which the present strategy for selecting control sets with guidance from quantum probability current
can be transferred to experimentally realizable settings and other
quantum transport networks.

\appendix

\section{Adjoint equation and GRAPE gradient}
\label{app:grape}

To evaluate the control gradient required by GRAPE, we introduce the
augmented functional
\begin{equation}
\eta'[\rho,\lambda,u]
=
\Tr{\bsP_s\rho(T)}
+
\int_0^T
\Tr{
\lambda^\dagger(t)
\left[
\dot{\rho}(t)-\mathcal L_{u(t)}\rho(t)
\right]
}\dif t ,
\label{eq:appendix-augmented}
\end{equation}
where \(\lambda(t)\) is a time-dependent operator Lagrange multiplier
that enforces the dynamical constraint.

Taking the variation with respect to \(\rho(t)\), integrating the
time-derivative term by parts, and using the adjoint Liouvillian
defined through
\begin{equation}
\Tr{\bsA^\dagger\mathcal L_{u(t)}(\bsB)}
=
\Tr{
\left[
\mathcal L_{u(t)}^\dagger(\bsA)
\right]^\dagger \bsB
},
\end{equation}
for arbitrary operators \(\bsA\) and \(\bsB\), gives
\begin{align}
\delta_\rho\eta'
&=
\Tr{
\left[
\bsP_s+\lambda(T)
\right]^\dagger
\delta\rho(T)
}+
\int_0^T
\Tr{
\left[
-\dot{\lambda}(t)
-\mathcal L_{u(t)}^\dagger\lambda(t)
\right]^\dagger
\delta\rho(t)
}\dif t .
\end{align}
Stationarity therefore yields the adjoint equation
\begin{equation}
\dot{\lambda}(t)
=
-\mathcal L_{u(t)}^\dagger\lambda(t),
\qquad
\lambda(T)=-\bsP_s.
\label{eq:appendix-adjoint}
\end{equation}
After this stationarity condition has been obtained, \(\lambda(t)\) is
identified as the adjoint variable, consistently with
Sec.~\ref{sec:model-dynamics}.

For the GRAPE discretization, the control terms are piecewise constant
on \(K\) intervals of duration \(\Delta t=T/K\). At GRAPE iteration
\(r\), let \(\mathcal L_k^{(r)}\) denote the Liouvillian on interval
\(k\). The state is propagated forward according to
\begin{equation}
\rho_{k+1}^{(r)}
=
\exp\left(
\mathcal L_k^{(r)}\Delta t
\right)\rho_k^{(r)},
\end{equation}
whereas the adjoint variable is propagated backward as
\begin{equation}
\lambda_k^{(r)}
=
\exp\left[
\left(\mathcal L_k^{(r)}\right)^\dagger\Delta t
\right]\lambda_{k+1}^{(r)},
\qquad
\lambda_K^{(r)}=-\bsP_s.
\label{eq:appendix-backward}
\end{equation}

The controlled Hamiltonian is written as
\begin{equation}
\bsH_u(t)
=
\bsH
+
\sum_{(m,n)\in\mathcal E}
u_{mn}(t)\bsH_{mn},
\qquad
\bsH_{mn}
=
\ket{m}\bra{n}+\ket{n}\bra{m},
\end{equation}
so that
\begin{equation}
\mathcal L_{u(t)}\rho
=
-\mathrm{i}[\bsH_u(t),\rho]
+
\mathcal L_{\mathrm{diss}}\rho.
\end{equation}
Since the dissipative part is independent of the control terms,
\begin{equation}
\frac{\partial\mathcal L_k^{(r)}}
{\partial u_{mn,k}^{(r)}}\rho
=
-\mathrm{i}
[\bsH_{mn},\rho].
\label{eq:appendix-control-liouvillian}
\end{equation}

Variation of Eq.~\eqref{eq:appendix-augmented} with respect to
\(u_{mn}(t)\), after imposing the forward and adjoint equations, gives
the GRAPE gradient. To first order in \(\Delta t\), its endpoint form is
\begin{align}
g_{mn,k}^{(r)}
&\equiv
\frac{\partial\eta^{(r)}}{\partial u_{mn,k}^{(r)}}
=
\operatorname{Re}\left\{
\mathrm{i}\Delta t\,
\Tr{
\left(\lambda_{k+1}^{(r)}\right)^\dagger
[\bsH_{mn},\rho_k^{(r)}]
}
\right\}.
\label{eq:appendix-endpoint-gradient}
\end{align}

In the numerical calculations, we instead evaluate the gradient at
the midpoint of each control interval. Consistently with
Sec.~\ref{sec:model-dynamics}, define
\begin{align}
\rho_{k+1/2}^{(r)}
&=
\exp\left(
\mathcal L_k^{(r)}\Delta t/2
\right)\rho_k^{(r)},\\
\lambda_{k+1/2}^{(r)}
&=
\exp\left[
\left(\mathcal L_k^{(r)}\right)^\dagger\Delta t/2
\right]\lambda_{k+1}^{(r)}.
\end{align}
the gradient used by GRAPE is
\begin{equation}
g_{mn,k}^{(r)}
\equiv
\frac{\partial\eta^{(r)}}{\partial u_{mn,k}^{(r)}}
=
-\Delta t\,
\operatorname{Re}
\left\langle
\lambda_{k+1/2}^{(r)},
\mathcal L_{mn}\rho_{k+1/2}^{(r)}
\right\rangle_{\mathrm{HS}}.
\label{eq:appendix-midpoint-gradient}
\end{equation}
Here, \(\mathcal L_{mn}\rho=-\mathrm{i}[\bsH_{mn},\rho]\), as in
Sec.~\ref{sec:model-dynamics}.

Thus, each GRAPE iteration consists of a forward propagation of
\(\rho^{(r)}\), a backward propagation of \(\lambda^{(r)}\), and
evaluation of Eq.~\eqref{eq:appendix-midpoint-gradient} for every
active controlled edge and time interval. For the projected GRAPE
stage, the normalized ascent direction is
\begin{equation}
d_{mn,k}^{(r)}
=
\frac{g_{mn,k}^{(r)}}
{\max_{(p,q)\in\mathcal A,\,\ell}
\left|g_{pq,\ell}^{(r)}\right|},
\qquad
(m,n)\in\mathcal A.
\end{equation}
To see why this direction can improve the transfer efficiency, we
regard the terminal efficiency as a function of the discretized
control vector \(\bsU\). For a control increment \(\delta\bsU\), its
first order variation is
\begin{equation}
\eta\!\left(\bsU^{(r)}+\delta\bsU\right)
-\eta\!\left(\bsU^{(r)}\right)
=
\sum_{(m,n)\in\mathcal A}\sum_{k=0}^{K-1}
g_{mn,k}^{(r)}\,\delta u_{mn,k}
+\mathcal O\!\left(\lVert\delta\bsU\rVert^2\right).
\label{eq:appendix-first-order-increase}
\end{equation}
Before imposing the amplitude bounds, the GRAPE step
\(\delta u_{mn,k}=\alpha_r d_{mn,k}^{(r)}\) therefore satisfies
\begin{equation}
\sum_{(m,n)\in\mathcal A}\sum_{k=0}^{K-1}
g_{mn,k}^{(r)}\,\delta u_{mn,k}
=
\frac{\alpha_r}
{\max_{(p,q)\in\mathcal A,\,\ell}
\left|g_{pq,\ell}^{(r)}\right|}
\sum_{(m,n)\in\mathcal A}\sum_{k=0}^{K-1}
\left|g_{mn,k}^{(r)}\right|^2
\geqslant 0.
\label{eq:appendix-grape-ascent}
\end{equation}
The inequality is strict whenever the gradient is nonzero. Hence, for
a sufficiently small positive step size, GRAPE changes the control
terms in a direction that increases \(\eta(T)\) to first order.
The trial and projected updates use the same symbols as in
Sec.~\ref{sec:model-dynamics},
\begin{align}
\widetilde u_{mn,k}^{(r+1)}
&=
u_{mn,k}^{(r)}+\alpha_r d_{mn,k}^{(r)},\\
u_{mn,k}^{(r+1)}
&=
\operatorname{clip}
\left(
\widetilde u_{mn,k}^{(r+1)},
-u_{\max},
u_{\max}
\right),
\label{eq:appendix-grape-update}
\end{align}
where \(\mathcal A\subseteq\mathcal E\) is the set of actively
optimized edges and \(\alpha_r\) is the step size. Control terms on
edges outside \(\mathcal A\) remain fixed at zero. If a component is
already at an amplitude bound, the projection may suppress an
outward component of the ascent step, but it does not reverse the
direction of any feasible component. The step size is reduced when
necessary, and a trial control is accepted only after forward
propagation confirms an actual increase in \(\eta(T)\). Consequently,
the accepted GRAPE iterations improve the transfer efficiency until
the gradient vanishes or the active amplitude bounds prevent further
ascent. This mechanism provides local improvement and does not imply
that the resulting control is a global optimum. The same analytic
gradient is subsequently supplied to the L-BFGS-B refinement.

\end{document}